\documentstyle[12pt]{article}
\newcommand{\bce}{\begin{center}}
\newcommand{\ece}{\end{center}}
\newcommand{\beq}{\begin{equation}}
\newcommand{\eeq}{\end{equation}}
\newcommand{\bea}{\vspace{0.25cm}\begin{eqnarray}}
\newcommand{\eea}{\end{eqnarray}}

\newcommand{\bsigma}{\mbox{\boldmath $\sigma$}}

\newcommand{\ba}{\begin{array}}
\newcommand{\ea}{\end{array}}

\def\lsim{\mathrel{\rlap{\lower4pt\hbox{\hskip1pt$\sim$}}
    \raise1pt\hbox{$<$}}}         %less than or approx. symbol
\def\gsim{\mathrel{\rlap{\lower4pt\hbox{\hskip1pt$\sim$}}
    \raise1pt\hbox{$>$}}}         %greater than or approx. symbol

\def\lsim{\mathrel{\rlap{\lower4pt\hbox{\hskip1pt$\sim$}}
    \raise1pt\hbox{$<$}}}         %less than or approx. symbol
\def\gsim{\mathrel{\rlap{\lower4pt\hbox{\hskip1pt$\sim$}}
    \raise1pt\hbox{$>$}}}         %greater than or approx. symbol

  \advance\oddsidemargin  by -1in
  \advance\evensidemargin by -1in
  \advance\topmargin      by -0.5in
\def\lsim{\mathrel{\rlap{\lower4pt\hbox{\hskip1pt$\sim$}}
    \raise1pt\hbox{$<$}}}         %less than or approx. symbol
\def\gsim{\mathrel{\rlap{\lower4pt\hbox{\hskip1pt$\sim$}}
    \raise1pt\hbox{$>$}}}         %greater than or approx. symbol

\def\beq{\begin{equation}}
\def\endeq{\end{equation}}
\def\arr{\begin{eqnarray}}
\def\endarr{\end{eqnarray}}
\makeindex

\begin{document}

\phantom.\hspace{8.8cm}{\large\bf KFA-IKP(TH)-1997-09\bigskip\\}
\phantom.\hspace{10.2cm}{\large \bf 6 May   1997}\vspace{1.5cm}\\
\begin{center}
{\Large \bf
Coherent neutrino magnetic conversion in crystals.\\

\vspace{1.0cm}}

{\large \bf
V.R.Zoller\medskip\\ } 
{\large \it
Institut  f\"ur Kernphysik, Forschungszentrum J\"ulich,\\
D-52425 J\"ulich, Germany {\footnote {KPH166@AIX.SP.KFA-JUELICH.DE}}\medskip\\
Institute for Theoretical and Experimental Physics,\\
ul. B.Chermushkinskaya 25, 117218 Moscow, Russia
{\footnote { ZOLLER@HERON.ITEP.RU}}\vspace{1cm}\\}
{\bf           Abstract}
\end{center}
We study coherent enhancement of neutrino electro-magnetic
conversion in crystals. Large coherence length which grows
with the neutrino energy makes the coherent enhancement
particularly effective for very small masses. We derive
constraints on the conversion rate which follow from the
Fresnel effects in scattering of neutrinos on atomic chains.
We comment on possible applications of the crystal converter
to searches for non-diagonal  neutrino magnetic transitions
in the CERN neutrino beam.

\newpage

In \cite{RARE} the  multiple scattering theory of Coulomb
coherent excitation of  high-energy particles propagating through
a crystal was  developed.
The coherent resonance enhancement of rare processes 
like the weak radiative transition
$p\gamma \rightarrow \Sigma^+$ \cite{P-SH} has been studied
in detail. In the plane wave Born approximation one would expect
the transition amplitude on a chain of $N$ identical atoms  
 ${\cal T}_{\Sigma^+ p}\propto N$ for 
the  beam momentum adjusted to satisfy the coherency condition at
a given lattice spacing $d$ \cite{DUB} 
(for earlier discussions on coherent
enhancement of atomic transitions see \cite{O3}).
The important finding of
Ref. \cite{RARE} is that distortions by the initial/final state
interactions of the proton/sigma with a Coulomb field of a crystal
suppress the coherent enhancement dramatically: ${\cal T}_
{\Sigma^+ p}\propto \log N$ already at $N\gsim (Z\alpha)^{-1}$.

Still another interesting process is the neutrino magnetic conversion
\cite{VVO} in a
Coulomb field of crystals
\beq
\nu_1\gamma\to \nu_2 \, .\label{eq:NUNU}
\endeq
Hereafter, $\nu_1$ and $\nu_2$ are the neutrino mass eigenstates.
In this case the initial/final state interaction effects are
negligible for all practical purposes and the law
\beq
{\cal T}_{21}=t_{21}N  \label{eq:SIMN}
\endeq
is expected to hold \cite {RARE}, where $t_{21}$
 is the transition amplitude on an isolated atom.
Recently there was much discussion on possible laboratory
observations of neutrino electro-magnetic conversion, for instance,
using the high-quality resonant
cavity converters \cite{GON,MAT,MAK}. In these experiments one deals with
 transition form factor at the photon virtuality 
 as small as $q^2\sim (10^{-6}\, eV)^2$. The bounds on the neutrino
magnetic moment which comes from the high-energy $\nu_{\mu} e$ elastic 
scattering correspond to quite different scale $q^2\sim (100\, MeV)^2$.
  A possibility
of the coherent enhancement  of neutrino transitions in
crystal converters must not be overlooked since it provides
the information on still another  region 
of $q^2\sim (1-10\, KeV)^2$.
Exploring this region of $q^2$ with crystal converters is topical in view of
a possible nontrivial $q^2$ 
 dependence,
resulting in a strong low-momentum enhancement of
 a neutrino magnetic form factor. A possible mechanism of such 
enhancement at very small $q^2$ was discussed in \cite{FRERE}.

 In this communication 
based on the multiple scattering theory 
we discuss the salient properties of crystal converters. We derive
the upper bound for the crystal thickness $ L=Nd$  at which the
conversion takes place on the crystal as a whole. This bound derives
from the Fresnel corrections to 
the conventional Glauber-Gribov approximation  \cite{GLAU,GRIB}.  
We conclude that crystal converters
in the current CERN neutrino beam can provide useful bounds for
the transition magnetic and electric dipole moments of the neutrino.

We start with the brief overview of formalism developed in \cite{RARE}.
The  amplitude of coherent transition (\ref{eq:NUNU}) on a  chain
built out of $ N$ 
 identical atoms  is written down as follows
\beq
{\cal T}_{21}({\bf q})=t_{21}({\bf q_{\perp}})
\langle N|\exp(i{\bf q}{\bf
r})|N\rangle
\endeq
where $|N>$ is the  ground-state wave function of the N-atomic chain.
 For our purposes it is
sufficient to use
 the uncorrelated wave function of  the Gaussian form (see \cite{RARE} for more
detail).
Notice that the inter-atomic distances   are large
compared to the
 Thomas-Fermi screening radius of the atom, $d\gg a_{TF}$.

 From the phenomenological transition  matrix element \cite{VVO} 
\beq
{\cal M}(\nu_1\gamma\to\nu_2)=i\bar u_{2}
\left(\mu_{21}+\gamma_5 d_{21}\right)
\sigma_{\mu\nu} q^{\nu}\varepsilon^{\mu}u_1 \label{eq:M}
\endeq
one readily finds the helicity-flip amplitude $t_{21}$
\beq
{t}_{21}({\bf q}_{\perp})
=Z\sqrt{\alpha}{\left[i\mu_{21}\bsigma\left[{\bf q}_{\perp}\times{\bf n}\right]
+d_{21}\bsigma{\bf q}_{\perp}\right]
\over q^2_{\perp}+\mu^2}          \,.                   \label{eq:T21}
\endeq
Here $\mu=a_{TF}^{-1}$ and   $a_{TF} =(m_{e}\alpha Z^{1/3})^{-1}$
is the Thomas-Fermi radius of an atom,
 ${\bsigma}$ is the Pauli spin vector and
${\bf n}$ -
 is a unit vector along the projectile direction. Our choice is  
${\bf n}=(0,0,1)$.
The non-diagonal magnetic and electric dipole moments are denoted by
$\mu_{21}$ and $d_{21}$, respectively.
 
 In terms of the  matrix element (\ref{eq:M})
the   cross section of the coherent neutrino conversion   
in crystal  
 is as follows 
\beq
{\sigma_{21} }= {\alpha Z^2}
{b^2_{21}}
\int d^2{ q}_{\perp}
{q_{\perp}^2\over
({q_{\perp}^2+\mu^2})^2}S^2_L(q_z)S_T^2(q_{\perp}) \label{eq:DIF}
\endeq
In eq.(\ref{eq:DIF})
$b^2_{21}=|\mu_{21}|^2+|d_{21}|^2$ for Dirac neutrinos and
$b^2_{21}=4\left[({\rm Im}\mu_{21})^2+({\rm Re} d_{21})^2\right]$
for Majorana neutrinos \cite{VVO}. 

Denoted by $S_T$ is the transverse form factor
\beq
S^2_T(q^2_{\perp})=\exp\left[-{2\over 3}\langle{\bf u}^2\rangle 
q^2_{\perp}\right]\,,
\endeq
where $\langle{\bf u}^2\rangle$ is the square of amplitude of the 
lattice thermal vibrations. It is precisely
the factor $S_T^2$ which cuts off  $q^2_{\perp}$ at $q^2_{\perp}\gsim 
{ \langle{\bf u}^2\rangle}^{-1} $.

The longitudinal form factor which involves the structure factor of a crystal
is defined as follows
\beq
S_L(q_z)=\exp\left[-{1\over 6}q_z^2\langle{\bf
u}^2\rangle\right]
{{\sin(q_z N_z d/2)}\over{\sin(q_z d/2)}}\,.   \label{eq:SQZ}
\endeq
The transition 
coherence length is determined by the inverse longitudinal momentum
transfer, $L_c\sim 1/q_z$.  In the very successful eikonal approach to
 the theory of multiple scattering on
extended targets \cite{GLAU,GRIB} it has been pointed out that
$q_z=\Delta m^2_{21}/2p$ \cite{GRIB}, where $\Delta m^2_{21}=m^2_2-m^2_1$.  
However, 
 the eikonal form of the wave function holds only at distances
from the scatterer shorter than the Rayleigh length (see \cite{GOTT}
and references therein)
\beq
 L_R=a^2_{TF}p\,.
\label{eq:RAY}
\endeq
Here $a_{TF}$ is the radius of the scatterer, which in our problem
is the Thomas-Fermi radius of an atom: $a_{TF} =(m_{e}\alpha
Z^{1/3})^{-1}$. If the coherence length is larger than the
Rayleigh length, the Fresnel corrections to geometrical optics
must be included and in the calculation of the converted wave the
eikonal Green's function
$$G_{eik}({\bf b};z)=i\delta({\bf b})
\theta(z)\exp({ipz})$$
must be substituted by the Fresnel-Green function \cite{GOTT}
$$G_F({\bf b};z)=
{p\theta(z)\over 2\pi z}
\exp\left({ip{\bf b}^2\over 2 z}\right)
\exp({ipz})\,. $$
 The Fresnel effects
are especially important in the high-energy magnetic scattering
since the amplitude $t_{21}({\bf q}_{\perp})$ vanishes at $q_{\perp}\to 0$
and the forward scattering dominance \cite{GLAU, GRIB} does not hold.
Let $q_{\perp}$ be the transverse momentum of the converted neutrino.
Then the converted wave calculated with the Fresnel-Green function
acquires  the additional Fresnel phase factor
$\exp[izq^2_{\perp}/2p]$. Hence
\beq
q_z ={\Delta m^2_{21}+q^2_{\perp}\over 2p} \label{eq:QZ}
\endeq
The small-angle scattering regime is assumed to hold,
  $\theta^2\simeq q^2_{\perp}/p^2\ll 1$.

The coherent 
conversion rate
\beq
R={\sigma_{21}\over d^2}
\endeq
 exhibits the resonance enhancement $R\propto N^2$
at the projectile  momentum
\beq
p\simeq p_n={(\Delta m^2_{21}+q^2_{\perp})d\over {4 \pi n}}\,,\, 
n=\pm 1,\pm 2,...
  \label{eq:PEQ}
\endeq
which satisfy the equation 
\beq
q_z d=2\pi n\,.     \label{eq:KAPPAD}
\endeq
 As a function of beam
momentum  $R$ looks
like a series of narrow  resonances with the width
\beq
 \Gamma_n={\sqrt{3}\over 2}
{(\Delta m^2_{21}+q^2_{\perp})d\over {\pi^2 n^2 N}}\,,\,~~~ n\neq 0 . 
                                                        \label{eq:GAMMAN}
\endeq

The smallness of the ratio
$
{\Gamma_n/ p_n}\sim { N^{-1}}
$
 leaves little hopes for the observation of these resonances in a wide band
neutrino beam. For the practical purposes, one must focus on the
coherent enhancement at $n=0$ in eq.(\ref{eq:KAPPAD}). 
In this case, ``the resonance width''
rises with the neutrino momentum. Indeed, for high energy
neutrino $q_z d$ is small and the 
longitudinal form factor in eq.(\ref{eq:DIF})
\beq
S_L^2\simeq N^2\left[1-
{N^2\over N^2_c}+...\right]\,.
\label{eq:SLSQ}
\endeq
defines the coherence length $L_c=N_cd$, where
\beq
N_c={4\sqrt{3}p\over (\Delta m^2_{21}+q^2_{\perp})d}\,. \label{eq:NCEIK}
\endeq
One finds the coherent enhancement $R \propto  N^2$
if $N\ll N_{c}$.

In contrast to the eikonal approximation, where
the coherence length is 
$L^{eik}_c\sim p/\Delta m^2_{21}$,
 the coherence length defined by eq.(\ref{eq:NCEIK}) depends strongly on  
the  transverse momentum $q_{\perp}$.
 The characteristic $q_{\perp}$ vary
in the range
$ 2\pi/d \lsim q_{\perp}\lsim a^{-1}_{TF},{\langle{\bf u}^2\rangle}^{-1/2}$ . 
 The thermal vibration amplitude
estimated from the Debye approximation
 corresponds to $\langle{\bf u}^2\rangle^{-1/2}\sim 20-40\, KeV$
for most commonly studied crystals at room temperature \cite{GEM}.
The inverse Thomas-Fermi radius  depends on $Z$ and  
is such that $a^{-1}_{TF}\sim 10-20\, KeV$. Typically, $2\pi/d\sim 1\, KeV$. 
These values of $q^2_{\perp}$ are much larger than 
the existing estimates of $\Delta m^2_{21}$  based on the observed
deficit of solar electron neutrinos, the deficit of atmospheric muon
neutrinos relative to electron neutrinos and a relative weight of hot
component in the dark matter of the Universe. They  vary from
$\Delta m^2_{21}\sim 10^{-6}\, eV^2$ to $\Delta m^2_{21}\sim 1\, eV^2$
( see \cite{MOH} for more discussion) and can safely be neglected
on the background of  lattice vibrations.
 
 As a function of crystal thickness  
 $R$ exhibits fast growth  with $L$ which is followed by the saturation regime
where  
$R\simeq 16\alpha Z^2{b^2_{21}p^2/ \mu^4d^4}$. Since we are interested in $L$ 
which are not much larger than $L_c$,
 the contribution from the incoherent transitions which 
is $\propto N$ 
as well as  corrections $\propto \log N$ at $N\gg N_c$ are neglected.

 The above peculiarities of $R$ 
are illustrated by Fig.1 in terms of the non-diagonal
transition magnetic moment. 
We consider the $W$-crystal converter. The attainable sensitivity 
of the  CERN neutrino detectors on the measurement of 
the neutrino conversion rate
in 
the "disappearance type"
experiment is assumed to be  $\simeq 10^{-4}$ \cite{GON}.
Somewhat deliberately we put $|\mu_{21}|=|d_{21}|$.
 So far as $L\ll L_c$ the conversion rate follows the geometrical
 optics. The upper bound on the transition magnetic moment
also follows the bound given by the geometrical optics.
Fresnel effects enter the game at $L\sim L_c$. 
The variation of the slope
in the transitional region is due to the smallness of 
$\langle{\bf u}^2\rangle$ compared to $a^2_{TF}$
which is specific for the $W$-crystal \cite{GEM}. 
At still larger $L$ the ratio ${\mu_{21}/ \mu_{B}}$ is saturated
at the value
\beq
{\mu_{21}\over \mu_{B}}\simeq
{1\over Z}{m_e\over 2p}{d^2\over a^2_{TF}}\sqrt{{R\over 2\alpha}}\,. 
 \label{eq:RATIO}
\endeq
Here $\mu_B$ is the Bohr magneton.
The above estimate, eq.(\ref{eq:RATIO}), does not involve 
$\langle{\bf u}^2\rangle$ since $\langle{\bf u}^2\rangle\ll a^2_{TF}$. 
If, on the contrary, $\langle{\bf u}^2\rangle\gsim a^2_{TF}$ 
the value of ${\mu_{21}/ \mu_{B}}$ in a 
 saturation regime is determined by the effect of thermal vibrations.

In the above analysis the neutrino has been assumed 
 to    propagate
 along the 
 crystallographic axis. 
In optimization of the conversion rate one must be aware of 
the natural angular divergence of the neutrino beam.
 If the projectile momentum ${\bf p}$ has non-zero
 transverse components 
and  $\theta$
is the divergence of the beam with respect to a crystal alignment axis, then
the angular aperture $\theta_{coh}$  for which the coherency condition holds
can be estimated as
\beq
\theta_{coh}\lsim {\pi\over q_{\perp}dN}\sim {1\over N}\,. \label{eq:THETA}
\endeq
To an extent that the neutrino source can be treated as point-like,
 the angular divergence 
can be corrected for by proper 
alignment of elements of the multi-crystal converter.

{\bf Acknowledgments:}  

 Thanks are due to N.N. Nikolaev for
careful reading the manuscript and useful comments.
 Discussions with A.N. Rozanov and M.I. Vysotsky
 are greatfully acknowledged.\\

\newpage

{\bf FIGURE CAPTIONS}\\

{\bf Fig.1}\\
The upper bound on the non-diagonal
 neutrino magnetic moment in units of the Bohr magneton
 as a function of
 the $W$-converter thickness  is shown for 
different neutrino beam momenta.

\newpage\

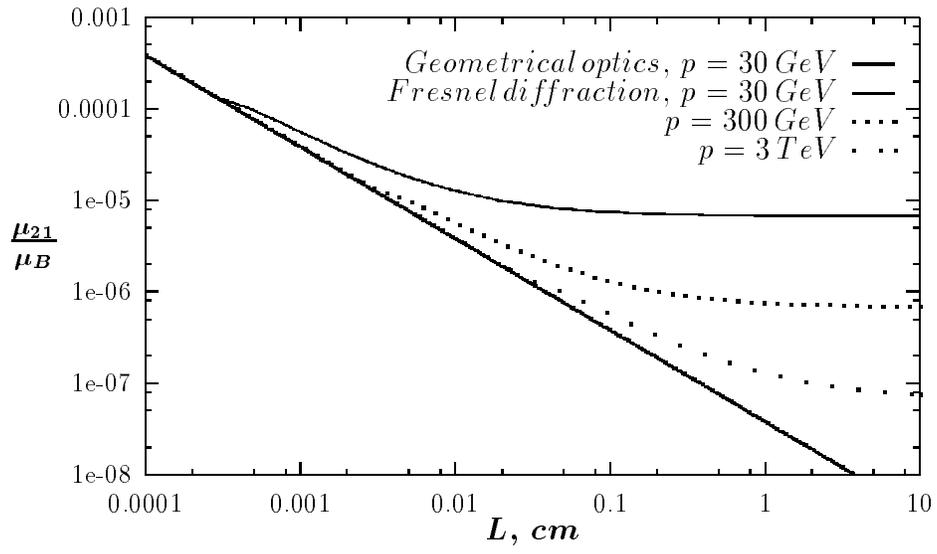
\end{document}